\documentclass{INTERSPEECH2023}

\interspeechcameraready

\usepackage{subcaption}
\usepackage{multirow}

\title{C\MakeLowercase{o}BERT: Self-Supervised Speech Representation Learning Through Code Representation Learning}
\name{Chutong Meng$^{1,\dag}$\thanks{$\dag$ Equal contribution. Work is done during internship at Bytedance.}, Junyi Ao$^{2,\dag}$, Tom Ko$^{1\ast}$\thanks{$\ast$ Corresponding author.}, Mingxuan Wang$^{1}$, Haizhou Li$^{2}$}
\address{
  $^1$ByteDance \\
  $^2$Shenzhen Research Institute of Big Data, School of Data Science, The Chinese University of Hong Kong, Shenzhen, China
}
\email{
mengct00@gmail.com,
junyiao1@link.cuhk.edu.cn,
\{tom.ko, wangmingxuan.89\}@bytedance.com,
haizhouli@cuhk.edu.cn
}

\begin{document}

\maketitle
 
\begin{abstract}
Speech is the surface form of a finite set of phonetic units, which can be represented by discrete codes.
We propose the \textbf{Co}de \textbf{BERT} (CoBERT) approach for self-supervised speech representation learning.
The idea is to convert an utterance to a sequence of discrete codes, and perform code representation learning, where we predict the code representations based on a masked view of the original speech input.
Unlike the prior self-distillation approaches of which the teacher and the student are of the same modality, our target model predicts representations from a different modality.
CoBERT outperforms the most recent state-of-the-art performance on the ASR task and brings significant improvements on the SUPERB speech translation (ST) task.
% Our code and models are released at \url{https://anonymous.4open.science/r/CoBERT-2023/}.
Our code and models are released at \url{https://github.com/mct10/CoBERT}.
\end{abstract}

\section{Introduction}

Self-supervised speech representation learning has become an important research direction after the success of BERT \cite{devlin2018bert} in natural language processing.
A key motivation is that effective speech representations simplify the downstream tasks, thus, reducing the required amount of annotated data for supervised fine-tuning. 
Along this idea, a number of  training approaches have been studied~\cite{van2018representation,schneider2019wav2vec,chung2020generative,hsu2021hubert,chen2021wavlm,baevski2022data2vec,cheng2022m3st,ma2022mt4ssl}, that include wav2vec \cite{Baevski2020vq-wav2vec,baevski2020discretebert,baevski2020wav2vec}, HuBERT \cite{hsu2021hubert}, and the recent data2vec \cite{baevski2022data2vec}.

While unpaired speech data are exploited in self-supervised learning, unpaired text data~\cite{ao-etal-2022-speecht5,bapna2021slam,tang-etal-2022-unified} are found to be useful as well.
The studies on speech-text pre-training  have shown
promising results, which seek to derive unified representations through joint training of between speech and text.
These methods can be seen as an attempt to reduce the gap between representations of speech and text.
A common problem for these methods is that it is difficult to align the representations of unpaired speech and text.
This problem can be alleviated by using paired speech recognition data \cite{ye-etal-2022-cross}.

In this paper, we propose the \textbf{Co}de \textbf{BERT} (CoBERT) approach for self-supervised speech representation learning.
Our method is inspired by the recent success of speech-text pre-training approaches and the data2vec approach.
First, we believe that the performance gain by speech-text pre-training arises from the improved speech encoder which has captured extra information from the text representations. 
Although it is difficult to obtain all the transcripts for the unpaired speech data, recent studies \cite{ao22_interspeech} show that self-organized codes are closely related to the text of the original speech.
Second, data2vec encourages the way of predicting contextualized latent representations in a self-distillation setup.
Thus, we examine code representation learning to benefit speech representation learning.
The core idea of our approach is to convert the speech to a sequence discrete units (codes), and perform code representation learning, where we predict the code representations based on a masked view of the original speech input. 
As the codes are generated from the unpaired speech data, the alignment problem encountered by speech-text pre-training approaches can be avoided.

The contributions of this work include, (i) We explore code representation learning using a self-distillation approach to benefit speech representation learning.
(ii) CoBERT can successfully predict contextualized latent representations from different and multiple modalities at the same time.
(iii) Experiments show that CoBERT relatively reduces the word error rate (WER) by 6.7\% on the LibriSpeech subsets \cite{panayotov2015librispeech} and brings significant improvements on the SUPERB ST task \cite{tsai-etal-2022-superb} by around 1.35 BLEU compared to data2vec.

The rest of this paper is organized as follows. In Section 2, we discuss the related work. Section 3 illustrates our proposed approach, including code generation, code representation learning and speech representation learning.
We share the experimental setting, results, and analyses in Section 4. 
Section 5 concludes the study.

\begin{figure*}[htb]
\begin{minipage}[b]{.59\linewidth}
  \centering
  \centerline{\includegraphics[width=\linewidth]{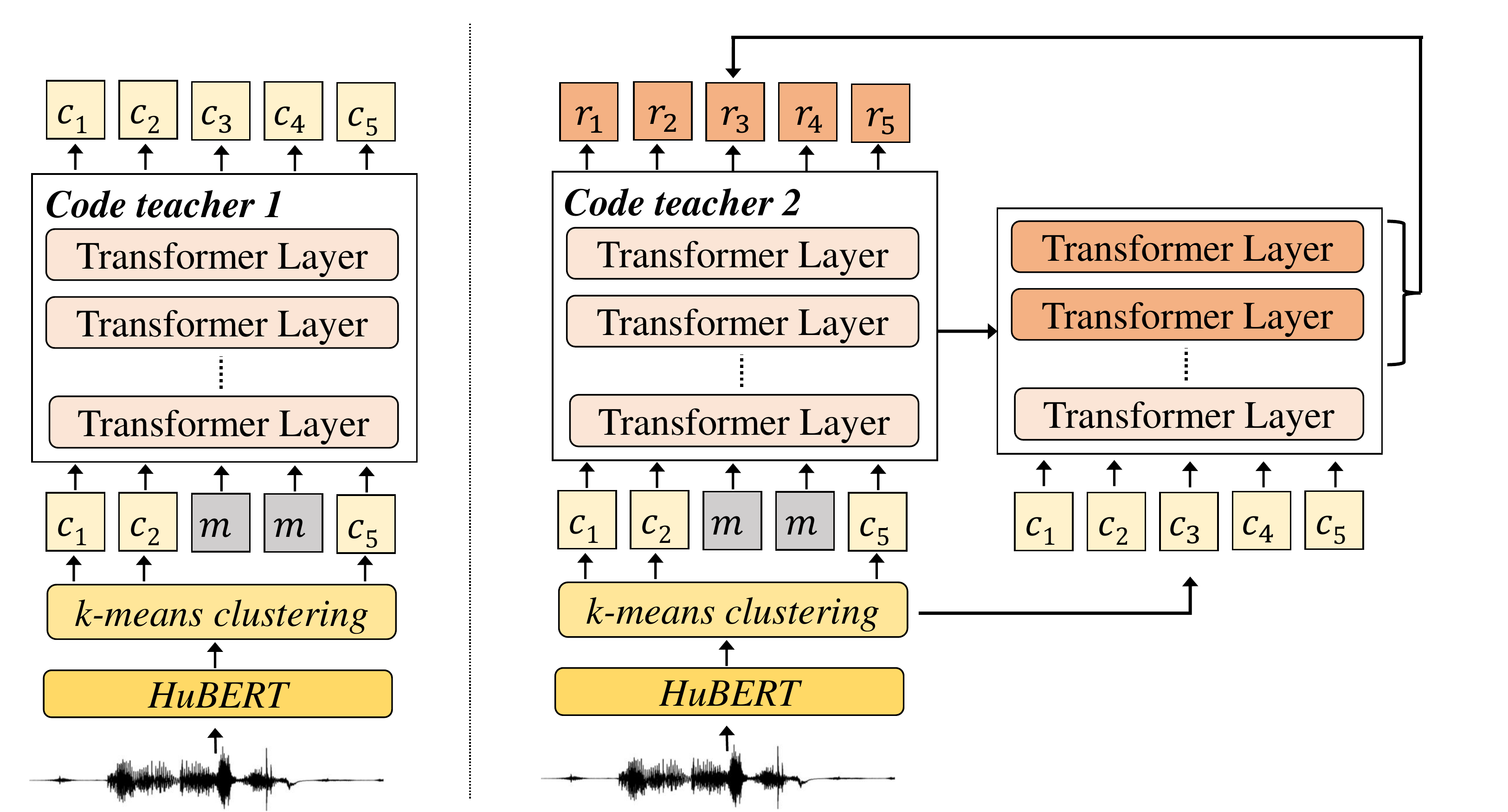}}
%  \vspace{1.5cm}
  \subcaption{}\medskip
\end{minipage}
\hfill
\begin{minipage}[b]{0.40\linewidth}
  \centering
  \centerline{\includegraphics[width=\linewidth]{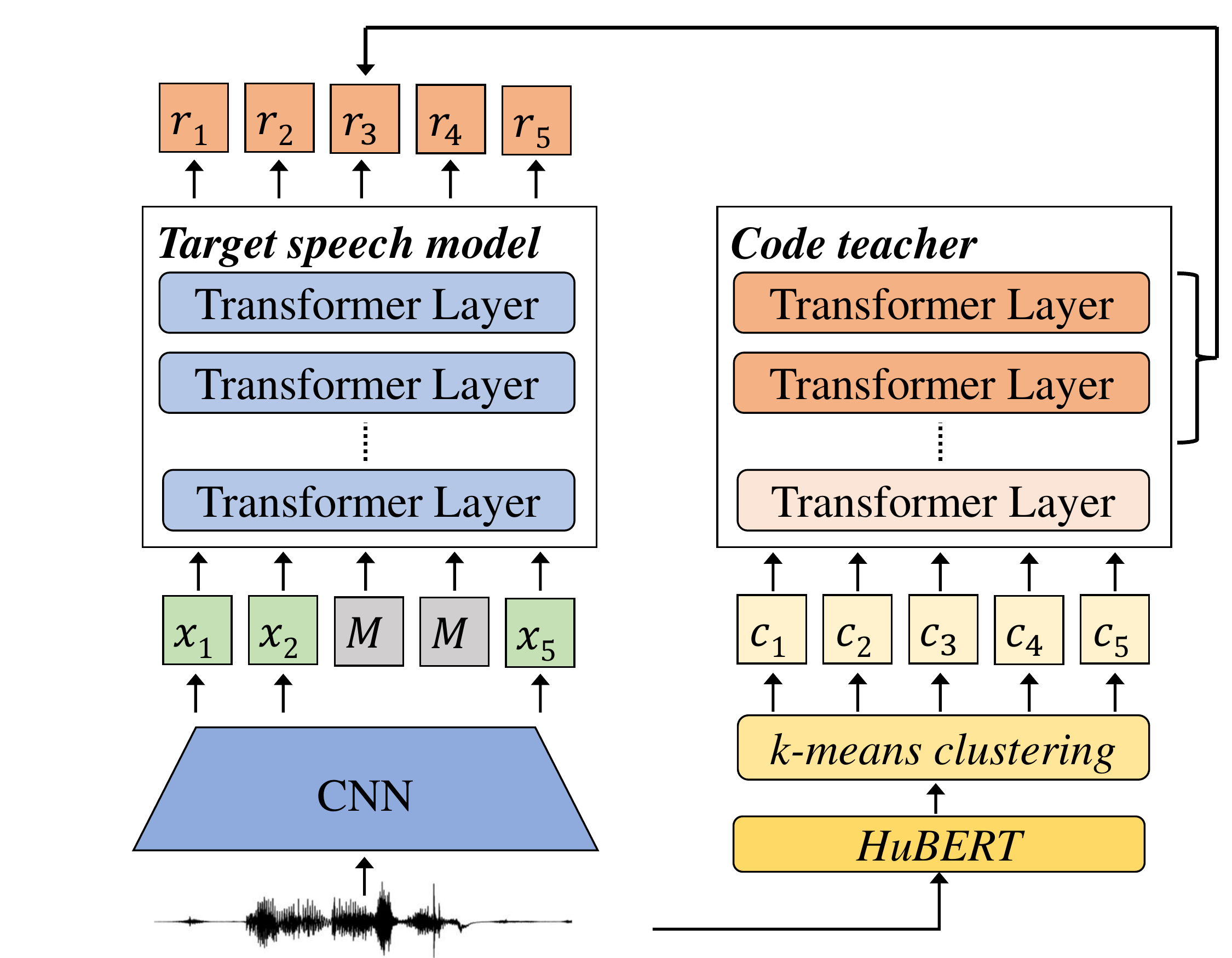}}
%  \vspace{1.5cm}
  \subcaption{}\medskip
\end{minipage}
\caption{Illustration of CoBERT.
The input codes for the code teachers are obtained by k-means clustering on HuBERT features. 
(a) We examine two code representation learning methods.
  The left one is based on BERT and the right one is based on data2vec.
(b) The representations produced by the code teacher are regressed by the target speech encoder given a masked view of the speech input.}
\label{fig:CoBERT}
\vspace{-12pt}
\end{figure*}

\section{Related work}

CoBERT is motivated by and most related to vq-wav2vec \cite{Baevski2020vq-wav2vec}, HuBERT \cite{hsu2021hubert} and data2vec \cite{baevski2022data2vec}.
vq-wav2vec is a pioneer work in exploring code representation learning. 
Further, they show that BERT-based representation learning is more effective in discrete than in continuous signal spaces \cite{baevski2020discretebert}.
HuBERT proposes to generate codes with an offline clustering step. 
Then it performs a BERT-based pre-training using these codes as target labels.
data2vec outperforms BERT-based pre-training methods by promoting a direct prediction of contextualized latent representations instead of modality-specific targets.
Our method integrates techniques of these methods.

We follow HuBERT's clustering procedures to obtain the codes so that CoBERT can perform representation learning completely in discrete code space. 
In comparison to data2vec, we follow their distillation approach to transfer the knowledge learnt from the code to the speech encoder, but our method differs in that CoBERT predicts representations of a different modality.
The teacher is a code encoder but the student is a speech encoder.
Moreover, CoBERT obtains better results when it predicts two sets of representations, one from the speech and one from the code, at the same time.

Another line of research aims at utilizing unpaired text to benefit speech representation learning \cite{ao-etal-2022-speecht5,bapna2021slam,tang-etal-2022-unified}. 
In this paper, we do not compare our results with these methods as we pre-train models only with unpaired speech.

\section{Method}
Our approach (Fig. \ref{fig:CoBERT}) can be divided into three steps: code generation, code representation learning and knowledge distillation.

\subsection{Code generation}

We follow the hidden units discovery steps described in \cite{hsu2021hubert} to extract the codes.
More precisely, we feed the raw waveform into a HuBERT model, and then apply k-means clustering to the latent features at a certain transformer layer. The generated codes are denoted as $C=[c_1,...,c_T]$ where $T$ is the number of frames, $c_t\in[K]$, and $K$ is the number of k-means classes.

We obtain codes from the 6\textsuperscript{th} layer features of HuBERT \textsc{Base}-it1 as well as 9\textsuperscript{th} layer features of HuBERT \textsc{Base}-it2. 
We train HuBERT \textsc{Base}-it1 ourselves, with the raw waveform as inputs and k-means clustering results on Mel-frequency cepstral coefficients (MFCC) features as targets.
We use the \textsc{Base} model released by Fairseq {\footnote[1]{https://dl.fbaipublicfiles.com/hubert/hubert\_base\_ls960.pt}} as the HuBERT \textsc{Base}-it2 model.
As shown in Table \ref{exp_codequality}, codes from HuBERT \textsc{Base} it2-L9 have better quality, so we use them in the subsequent steps.

\subsection{Code representation learning}

We examine code representation learning in two different ways (Fig. 1(a)).
%two methods to train code teachers and learn code representations.

Code teacher 1 is based on mask language modeling (MLM) similar to BERT \cite{devlin2018bert}. 
Let $Z$ denote the indices where the elements are to be masked. Also let $\widetilde{C}=mask(C, Z)$, which replaces the codes in $C$ whose indices lie in $Z$ with a special code $M$. Then the model is trained to predict the masked codes. We only compute the loss over the masked positions, which is

\begin{equation}
    L_{mlm}=-\sum_{t\in Z}{\log{p(c_t|\widetilde{C}, t)}}.
\end{equation}
We use the same HuBERT \textsc{Base} model as in \cite{hsu2021hubert}, except that we replace the waveform encoder with a word embedding layer.

Code teacher 2 is based on the self-distillation approach similar to data2vec \cite{baevski2022data2vec}.
%There is a teacher model and a student model. 
The teacher model takes $C$ as input and produces target representations $R=[r_1,...,r_T]=\frac{1}{L}\sum_{l=N-L+1}^{N}{\hat{a}^l}$, where $N$ is the number of transformer layers and features from each of the top $L$ layers will be normalized to $\hat{a}^l$ then summed to targets. The student model takes in $\widetilde{C}$ and regresses $R$. 

The parameters of the teacher model are exponentially moving average (EMA) of those of the student model: $\theta_t \leftarrow \tau\theta_t + (1 - \tau)\theta_s$, where $\theta_t$ is the parameters of the teacher model and $\theta_s$ is the parameters of the student model.
Only the student model is updated via backpropagation. The objective is an $L2$ loss,

\begin{equation}
    L_{sd}=\frac{1}{2}(\hat{R}-R)^2.
\end{equation}
where $\hat{R}$ is the predictions of the student model and $sd$ denotes self-distillation. 

For code teacher 2, we apply the data2vec \textsc{Base} model with its waveform encoder replaced with a word embedding layer.
We use the features from the top 8 layers to generate targets, i.e., $L=8$.
We adopt slightly different masking strategies for the two teachers. Let $p$ denote the probability of an input element chosen as the start position of a mask span.
We adopt $p=8\%$ for MLM based training and $p=6.5\%$ for self-distillation based training.
For both models, we set their mask spans to 10. 

\begin{table*}[!ht]
\centering
\caption{\label{table:main_results} Main results on ASR and ST tasks. The WER scores are evaluated on the test-clean and test-other sets of LibriSpeech when using the 10 hours or 100 hours subsets as the training data and the BLEU scores are evaluated on the test set of the CoVoST2 En $\rightarrow$ De dataset. The WER scores of HuBERT and data2vec without LM and the BLEU score of data2vec are obtained by fine-tuning the released model as they are not reported in the corresponding papers.}

\resizebox{\textwidth}{!}{
\begin{tabular}{@{\extracolsep{3pt}}lccccccccccc}
\toprule
\multirow{3}{*}{Model} & \multicolumn{4}{c}{\textbf{10 hours subset}} & \multicolumn{4}{c}{\textbf{100 hours subset}} & \multirow{3}{*}{\textbf{SUPERB ASR}} & \multirow{3}{*}{\textbf{SUPERB ST}} \\
\cline{2-5} \cline{6-9} &  \multicolumn{2}{c}{No LM} & \multicolumn{2}{c}{4-gram} & \multicolumn{2}{c}{No LM} & \multicolumn{2}{c}{4-gram} &  \\
\cline{2-3} \cline{4-5} \cline{6-7} \cline{8-9} & clean & other & clean & other & clean & other & clean & other  \\
\midrule
\midrule
% \midrule
\textbf{\textit{Baselines}} \\
\midrule
DiscreteBERT \cite{baevski2020discretebert} & - & -  & 5.9 & 14.1 & - & -  & 4.5 & 12.1 & - & - \\
wav2vec 2.0 \textsc{Base} \cite{baevski2020wav2vec} & 11.1 & 17.6  & 4.3 & 9.5 & 6.1 & 13.3  & 3.4 & 8.0 & 6.43 & 14.81 \\
HuBERT \textsc{Base}  \cite{hsu2021hubert} & 10.1 & 16.8 & 4.3 & 9.4 & 5.6 & 12.7 & 3.4 & 8.1 & 6.42 & 15.53 \\
WavLM \textsc{Base}  \cite{chen2021wavlm} & 9.8 & 16.0 & 4.3 & 9.2 & 5.7 & 12.0 & 3.4 & 7.7 & 6.21 & - \\
data2vec \textsc{Base} \cite{baevski2022data2vec} & 7.2 & 12.3 & 3.9 & 8.1  & 4.2 & 9.7 & \textbf{2.8} & 6.8 & 4.94 & 17.56 \\
\midrule
\textbf{\textit{Our Methods}} \\
\midrule
code teacher 1 (BERT)  & 8.8 & 14.5 & 5.1 & 9.5 & 5.0 & 10.8 & 3.6 & 8.1 & - & 17.11 \\
code teacher 2 (data2vec)  & 7.9 & 12.9 & 4.4 & 8.8 & 4.6 & 9.9  & 3.4 & 7.5 & - & 17.75 \\
\midrule
CoBERT & & & &  & && &  &   \\
- distilling code teacher 1 & 7.4 & 12.7 & 3.7 & 7.9 & 4.3 & 9.8  & 2.9 & 6.8 & - & 17.81 \\
\quad + self-distillation & 7.2 & 12.2 & 3.7 & 7.7 & 4.3 & 9.6 & 3.0 & 6.6 & - & 18.12 \\
- distilling code teacher 2 & 7.5 & 12.3 & 4.0 & 8.0 & 4.2 & 9.3  & 3.0 & 6.7 & - & 18.73 \\
\quad + self-distillation         & \textbf{6.8} & \textbf{11.4} & \textbf{3.6} & \textbf{7.4} & \textbf{4.0} & \textbf{8.9} & \textbf{2.8} & \textbf{6.4} & \textbf{4.74} & \textbf{19.10} \\
\bottomrule
\end{tabular}
}
\vspace{-10pt}
\end{table*}

\subsection{Knowledge distillation}

%data2vec is trained by predicting the model representations
%of the full input data given a partial view of the input 
The target speech encoder, CoBERT, is trained by predicting the representations of the code input given a masked view of the speech input (Fig. 1(b)). In other words, we distil the code model into a speech model.
%After pre-training code teachers, we teach a speech encoder by regressing the code representations based on a masked view of the speech.

In this paper, the architecture of the CoBERT model follows exactly the speech data2vec \textsc{Base} model \cite{baevski2022data2vec}.
It takes raw waveform as input.
Let $X=[x_1,\ldots,x_T]$ denote the speech hidden states after the waveform is downsampled by the same waveform encoder as the one described in \cite{baevski2022data2vec}.
We mask $X$ with $p=6.5\%$ and mask span 10 to generate a masked view $\widetilde{X}=mask(X,Z)$.
The pre-trained code teacher produces code representations $R_{code}$ based on the aligned code sequence corresponding to $X$.
%The code sequence corresponding to $X$ is fed to a code teacher, which will produce code representations $R_{code}$.
The objective is to regress the code representations: 
\begin{equation}
    L_{code}=\frac{1}{2}(R_{Co}-R_{code})^2
\end{equation}
where $R_{Co}$ is the representations generated by the CoBERT model.

Additionally, to fully utilize the speech information, our distillation process can be combined with the self-distillation setup used in data2vec. The teacher model, which averages the parameters of the CoBERT model exponentially, takes $X$ as input and produces speech representations $R_{speech}$ from its top 8 layers.
We use a separate linear layer on top of CoBERT model, which generates $R_{Co'}$, to regress $R_{speech}$.
The loss from the self-distillation teacher is therefore 
\begin{equation}
    L_{speech}=\frac{1}{2}(R_{Co'}-R_{speech})^2.
\end{equation}

%Note that there will be a separate projection layer on top of the student model for each teacher, so $R_{Co'}$ and $R_{Co}$ are different in order to learn representations from different modalities.

The final loss for CoBERT is a weighted sum of $L_{code}$ and $L_{speech}$:

\begin{equation}
    L_{CoBERT}=\alpha L_{code} + (1-\alpha)L_{speech}.
\end{equation}
where $0 \leq \alpha \leq 1$.
In all of our experiments, we set $\alpha=0.5$.

% \section{Results}
% \section{Discussion}

\section{Experiments}

\label{sec:majhead}
\subsection{Experiment Setup}
We conduct our experiments using Fairseq{\footnote[1]{https://github.com/pytorch/fairseq}} \cite{ott2019fairseq}.
For pre-training, we mainly follow the training process and hyper-parameters in \cite{hsu2021hubert,baevski2022data2vec}.
We use the full 960 hours training set of LibriSpeech \cite{panayotov2015librispeech} without the transcription for pre-training.
For code teacher 1, we optimize the model with Adam \cite{kingma2014adam} by warming up the learning rate for the first 8\% of updates to a peak of $5 \times 10 ^{-4}$, which is linearly decayed for the following updates.
For code teacher 2 and CoBERT experiments, we use Adam \cite{kingma2014adam} to optimize the model with a tri-stage scheduler, which linearly warms up the learning rate for the first 3\% of updates to a peak of $5 \times 10 ^{-4}$, holds it for 90\% of updates and linearly decays the learning rate for the following updates.
We pre-train the model for 400k updates on 16 GPUs with a batch size of around 10k tokens, 11.85k tokens and 237s samples per GPU for code teacher 1, code teacher 2 and CoBERT experiments.

For the ASR fine-tuning, we utilize 10 and 100 hours subsets, and segment the text with the character set.
The learning rate is warmed up for the first 10\% steps, held as a constant for the following 40\% steps, and decayed linearly for the rest steps.
For the 10/100 hours subset, we train the model with a learning rate of 5e-5/3e-5 for 20k/80k. 
For the 10 hours subset, the encoder part is fixed for the first 10k steps.

We evaluate our model on the ASR task by using wav2letter++ \cite{pratap2019wav2letter} beam search decoder with a beam size of 1500 for 4-gram language model-fused decoding, which optimized:

\begin{equation}
    \log P_{ctc}(Y|X) + \omega_1 \log P_{LM}(Y) + \omega_2 |Y|
\end{equation}
where $Y$ is the prediction of a text sequence, $|Y|$ is the sequence length, and $\omega_1$ and $\omega_2$ are the weights of the language model and word score, respectively.
The decoding hyperparameters, including language model weight, word score and silence weight, are searched with Ax\footnote{https://github.com/facebook/Ax}, which is a Bayesian optimization toolkit.
We use the official 4-gram language model \footnote{https://www.openslr.org/resources/11/4-gram.arpa.gz} trained on the LibriSpeech-LM corpus for inference.

For the SUPERB ASR and ST fine-tuning, the training and inference processes follow \cite{tsai-etal-2022-superb}. 
We use LibriSpeech and CoVoST2 en$\rightarrow$de dataset \cite{wang2020covost} for evaluation.
The downstream model for ST has three transformer layers for both the encoder and decoder, with an additional convolution layer for down-sampling the input.
The downstream model for ASR has two layers of BLSTM and is optimized by CTC loss on characters level.

\subsection{Main Results}
\label{ssec:main_results}

Table \ref{table:main_results} shows the main results of code teachers and CoBERT on ASR and ST tasks. 
We evaluate the WER scores on the standard LibriSpeech test-clean and test-other sets and BLEU scores on the test set of the CoVoST2 En $\rightarrow$ De dataset.
We compare our method with several works from the literature, including DiscreteBERT \cite{baevski2020discretebert}, wav2vec 2.0 \cite{baevski2020wav2vec}, HuBERT \cite{hsu2021hubert}, WavLM \cite{chen2021wavlm} and data2vec \cite{baevski2022data2vec}, which are competitive self-supervised approaches.
As the WER scores without LM of HuBERT and data2vec and the BLEU scores of data2vec are not reported in the corresponding papers, the results are obtained by fine-tuning the released model.

Experiments show that without LM fusion, code teacher 1 outperforms HuBERT on the ASR task, which relatively brings 12.1\% and 14.2\% reductions on the test-clean and test-other sets, while it performs worse or on par compared to the HuBERT baseline with LM fusion.
For the ST task, code teacher 1 achieves significant improvement by around 1.58 BLEU.
This indicates the code representations of code teacher 1 may carry extra semantic information.
In consideration with a similar observation in \cite{baevski2020discretebert}, we hypothesize that BERT-based representation learning may be more effective in discrete than in continuous signal spaces.

Without self-distillation, CoBERT models outperform DiscreteBERT, HuBERT and wav2vec 2.0 by a large margin for both the ASR and ST tasks, while performing better than data2vec on the ST task.
With self-distillation, CoBERT reaches a relatively 6.7\% WER reduction on the averages of all ASR sets and improves the BLEU score from 17.56 to 19.10 compared to the data2vec baseline.
\vspace{-5pt}
\subsection{Analysis}
\label{sssec:subsubhead4}

\subsubsection{Code quality across different features}
\label{sssec:subsubhead5}

As code quality may decide the final performance, we investigate the clustering quality across different features, including features of the 6th layer of HuBERT \textsc{Base}-it1, the 9th layer of HuBERT \textsc{Base}-it2, the 9th layer of data2vec, and the average top 8 layers of data2vec.
We generate frame-level phonetic transcripts using Montreal Forced Aligner \cite{mcauliffe17_interspeech} with the default pre-trained English acoustic model \cite{mfa_english_us_arpa_acoustic_2022} and dictionary \cite{gorman2011prosodylab}.
Then we follow \cite{hsu2021hubert} to compute the code quality metrics using the codes and the phonemes.
The results are shown in Table \ref{exp_codequality}.
Although the codes from it2-L9 are better than those from it1-L6, they achieve similar performance on the test-other subset of LibriSpeech. 
Therefore, it is fair for us to compare the performance of CoBERT and HuBERT.
We also explore whether we can derive better-quality codes from the data2vec model since it has a better ASR performance. 
However, it turns out that the qualities of codes from data2vec are much worse than HuBERT's. 

\vspace{-5pt}
\begin{table}[!ht]
\begin{center}
\caption{\label{exp_codequality} Quality of the clustering assignments and the effect on the ASR task with respect to different features.}

\resizebox{\linewidth}{!}{
\begin{tabular}{lccccc}
\toprule
model & feature & phone purity & cluster purity & PNMI & WER \\
\midrule
\midrule
\multirow{2}{*}{HuBERT}  & it1-L6  & 0.668 & 0.110 & 0.657 & 12.74 \\
& it2-L9 & 0.674 & 0.107 & 0.666 & 12.73 \\
\midrule
\multirow{2}{*}{data2vec}  & it1-L9 & 0.583 & 0.100 & 0.594 & - \\
 & Top8-avg & 0.536 & 0.111 & 0.540 & - \\
\bottomrule
\end{tabular}
}
\end{center}
\vspace{-30pt}
\end{table}

\subsubsection{Effect of different teachers and model input}
\label{sssec:teacher_ana}

In this section, we verify the importance of the code teachers by using different teachers for the distillation.
In table \ref{exp_differentteacher}, the parameters of the HuBERT model and code teachers are fixed throughout the distillation.
We intentionally not to further distil the data2vec checkpoint in table \ref{table:main_results} as it is already trained with self-distillation.

Firstly, we want to show that distilling code models into speech models is better than distilling speech models. 
In the experiments of using one teacher, distilling HuBERT performs better than HuBERT itself and distilling any of the code teachers achieves a significant improvement compared to it.
This indicates code teachers may contain extra language information compared to the conventional self-supervised speech models.

Secondly, we observe that distilling code teachers complements self-distillation and can achieve further improvement.
When self-distillation is applied, distilling code teacher 1 outperforms distilling HuBERT by 0.2/0.8 WER reduction on the ASR task and 0.98 BLEU improvement on the ST task.
This proves that improvements achieved by our method can not be achieved by simply distilling speech models.

\vspace{-5pt}
\begin{table}[!ht]
\begin{center}
\caption{\label{exp_differentteacher} Comparison of using different teacher models in terms of BLEU and WER scores. 
For ASR, the models are fine-tuned with 100 hours subset of LibriSpeech.}

\resizebox{\linewidth}{!}{
\begin{tabular}{@{\extracolsep{3pt}}lcccc}
\toprule
\multirow{2}{*}{Teacher model} &  \multicolumn{1}{c}{\textbf{BLEU} $\uparrow$}  & \multicolumn{2}{c}{\textbf{WER} $\downarrow$}  \\
\cline{2-2}\cline{3-4} &  SUPERB ST & test-clean & test-other \\
\midrule
\midrule
\textit{\textbf{W/o self-distillation}} \\
\midrule
HuBERT  & 17.19 & 5.0 & 10.6  \\
code teacher 1  & 17.81 & 4.3 & 9.8  \\
code teacher 2  & 18.73 & 4.2 & 9.3 \\
\midrule
\textit{\textbf{With self-distillation}} \\
\midrule
HuBERT  &  17.14 & 4.5 & 10.4 \\
code teacher 1  & 18.12 & 4.3 & 9.6 \\
code teacher 2  & 19.10 & 4.0 & 8.9  \\
\bottomrule
\end{tabular}
}
\end{center}
\vspace{-25pt}
\end{table}

\section{Conclusions}
We present CoBERT, a self-supervised speech representation learning approach which enables a speech student encoder to learn contextualized latent representations from both speech and code modalities.
First of all, we pre-train code models with HuBERT codes to benefit from training in discrete space.
Then, we distil the code model into a speech model, which aims at performing a better learning across modalities.
The significant improvement on the ST task indicates the representation of CoBERT may carry more language information compared to prior work.

In this paper, only codes generated from the unpaired speech are used to pre-train the code teacher and the amount of them is far less than the data scale to train a state-of-the-art text encoder. 
Future work may investigate the way to exploit text data to compensate for the paired code scarcity. 
We would also like to evaluate CoBERT on more spoken language understanding tasks and pre-train a multilingual CoBERT for multilingual speech translation task.

% \lipsum[66]

\section{Acknowledgements}
This work is supported by National Natural Science Foundation of China (Grant No. 62271432), Internal Project Fund from Shenzhen Research Institute of Big Data (Grant No. T00120220002), and Guangdong Provincial Key Laboratory of Big Data Computing, The Chinese University of Hong Kong, Shenzhen (Grant No. B10120210117-KP02).
% \ifinterspeechfinal
%      The INTERSPEECH 2023 organisers
% \else
%      The authors
% \fi
% would like to thank ISCA and the organising committees of past INTERSPEECH conferences for their help and for kindly providing the previous version of this template.

% As a final reminder, the 5th page is reserved exclusively for references. No other content must appear on the 5th page. Appendices, if any, must be within the first 4 pages. The references may start on an earlier page, if there is space.

\bibliographystyle{IEEEtran}
\bibliography{refs}

\end{document}